# DEVELOPING A MEASURE OF ONLINE WELLBEING AND USER TRUST

**Liz Dowthwaite, Elvira Perez Vallejos, Helen Creswick, Virginia Portillo, Menisha Patel, Jun Zhao**

University of Nottingham (UK), University of Nottingham (UK), University of Nottingham (UK), University of Nottingham (UK), University of Oxford (UK), University of Oxford (UK)

liz.dowthwaite@nottingham.ac.uk; elvira.perez@nottingham.ac.uk; helen.creswick@nottingham.ac.uk; virginia.portillo@nottingham.ac.uk; menisha.patel@cs.ox.ac.uk; jun.zhao@cs.ox.ac.uk

**ABSTRACT**

This paper describes the first stage of the ongoing development of two scales to measure online wellbeing and trust, based on the results of a series of workshops with younger and older adults. The first, the Online Wellbeing Scale includes subscales covering both psychological, or eudaimonic, wellbeing and subjective, or hedonic, wellbeing, as well as digital literacy and online activity; the overall aim is to understand how a user's online experiences affect their wellbeing. The second scale, the Trust Index includes three subscales covering the importance of trust to the user, trusting beliefs, and contextual factors; the aim for this scale is to examine trust in online algorithm-driven systems. The scales will be used together to aid researchers in understanding how trust (or lack of trust) relates to overall wellbeing online. They will also contribute to the development of a suite of tools for empowering users to negotiate issues of trust online, as well as in designing guidelines for the inclusion of trust considerations in the development of online algorithm-driven systems. The next step is to release the prototype scales developed as a result of this pilot in a large online study in to validate the measures.

**KEYWORDS:** wellbeing, trust, online experience, scale.

## 1. INTRODUCTION

As interaction with online platforms is becoming an essential part of people's everyday lives, the use of automated decision-making algorithms in filtering and distributing the vast quantities of information and content to users is having an increasing effect on society, with many people raising questions about the fairness, accuracy and reliability of such outcomes. Online users often do not know when to trust algorithmic processes and the platforms that use them, reporting anxiety and uncertainty, feelings of disempowerment, defeatism, and loss of faith in regulation (Creswick et al., 2019; Knowles & Hanson, 2018). Various other negative effects of using such online technologies have been identified, for example, concerns about hostile actors spreading online disinformation, vulnerable groups becoming victims of scams, harmful user-generated content and bullying, and addiction and excessive screen time (Chadborn et al., 2019; DCMS, 2019; Kidron et al., 2018; Livingstone et al., 2010). These issues lead to concerns about wellbeing which can affect both the user and broader society. It is therefore important that mechanisms and tools are developed to assess online wellbeing and trust with the view to support users interacting with the online world.

This paper describes the first stage of the ongoing development of an 'Online Wellbeing Scale' (OWS) and a 'Trust Index' (TI) to aid in understanding how trust (or lack of trust) relates to overall wellbeing online. There are two broad aims of the scales. For researchers, the scales will allow exploration of the relationship between wellbeing and trust, to understand how trust of algorithmic systems affects user's online experiences across different online activities and at different levels of digital literacy. For the users, the scales will contribute to the development of a tool for self-measuring and reflecting on trust, as part of engaging in dialogue with platforms in order to jointly recover from trust breakdowns. The scales will be part of a suite of tools for (1) empowering users to negotiate issues of trust online and (2) designing guidelines for the inclusion of trust relationships in the development of algorithm-driven systems.

## 2. BACKGROUND

Interaction with online platforms is becoming an essential part of people's everyday lives. As digital technology develops, people are using the Internet to carry out more and more of their everyday activities, from socialising and entertainment to financial transactions and working. News feeds, online media, search engine results and product recommendations increasingly use personalisation algorithms (i.e., sequences of instructions or commands for computers to solve a task) to help users cut through the vast amounts of available information. They use vast quantities of data, especially personal data, to provide a more personalised, appealing and engaging online experience. The use of such algorithms is therefore having an increasing effect on society, with many people raising questions about the fairness, accuracy and reliability of outcomes. Algorithmic decision-making often lacks transparency and when using online services users are generally given next-to-no information about the algorithms that are being used, or the data that is used to feed those algorithms. Online users therefore often do not know when to trust algorithmic processes and the platforms that use them, reporting anxiety and uncertainty, feelings of disempowerment, defeatism, and loss of faith in regulation (Creswick et al., 2019; Knowles & Hanson, 2018). These issues lead to concerns about wellbeing, which can affect both the user and broader society.

Other aspects of the online world have also raised concerns about wellbeing. Online media have been associated with both physical and mental harms (DCMS, 2019). This is particularly the case among children and young people; approximately 1 in 5 11-16 year-olds have been exposed to potentially harmful content online (Livingstone et al., 2010). There are also emerging challenges about designed addiction and excessive screen time (Kidron et al., 2018). An increase in time spent on social media has also been associated with decreased life satisfaction and quality of life in children (McDool et al., 2016), and may be damaging to mental health (Royal Society for Public Health, 2017). Vulnerable groups, including older adults, are often victims of financial scams or intimidation, and feel they have no choice but to access online services that they do not fully understand or trust (Chadborn et al., 2019). There are also serious concerns about hostile actors using online disinformation to undermine democratic values and principles. It is therefore important to be able to assess the effects of the online world on wellbeing to make meaningful recommendations for the responsible design of technologies. It has also been suggested that problematic internet and social media use may be linked to personality and psychological needs (Kozan et al., 2019). Whilst there are many existing measures of wellbeing, there are no measures of online wellbeing specifically, where 'online wellbeing' is defined as the effects of carrying out activities and tasks online on a person's wellbeing. This paper describes the first stages of creating such a scale, the Online Wellbeing Scale (OWS).

## 2.1. Measuring Online Wellbeing

There are many different definitions of wellbeing, often focusing on the different dimensions of wellbeing rather than a single concept (Dodge et al., 2012). Studies of wellbeing often fall into two main traditions: eudaimonic and hedonistic (Deci & Ryan, 2008). Eudaimonic wellbeing refers to living well and flourishing as a person, and is often conceptualised as psychological wellbeing (PWB), whilst hedonistic wellbeing refers to the balance of positive and negative emotions that are experienced by an individual, conceptualised as subjective wellbeing (SWB). The OWS will measure both types in order to get a broad idea of how the online world affects overall wellbeing. The scale will rely on self-report, as such measures allow individuals to define for themselves how they experience and understand their own wellbeing, rather than forcing an objective definition from outside (Alexandrova, 2005).

The attainment of PWB is part of the grounding of Self-Determination Theory (SDT) (Ryan et al., 2006). SDT is a collection of six sub-theories which present a framework for studying motivation and personality, focussing on how social and cultural factors affect people's agency, wellbeing, and performance. One of these sub-theories, Basic Psychological Needs Theory (BPNT) states that psychological health and wellbeing are achieved by satisfying certain basic needs: *autonomy* (i.e. agency, the freedom or independence to act as desired), *relatedness* (i.e. a social connection with others), and *competence* (i.e. self-efficacy, the ability to carry out an action effectively) (Deci & Ryan, 2000; Ryan & Deci, 2000, 2017). The three basic psychological needs are universal, having been found across many cultures (Chen et al., 2015) and domains including family, friends, relationships, school, work, and hobbies (Milyavskaya & Koestner, 2011). BPNT has been explored in a few online contexts, for example social media, where different activities were linked to a lack of particular needs (Masur et al., 2014), and online crowdsourcing where different gamification mechanisms satisfied needs to greater or lesser degrees (Goh et al., 2017). It has also been suggested that consideration of basic psychological needs is vital to improving the design of user experience (Peters et al., 2018; Wang et al., 2019). The Basic Psychological Need Satisfaction (BPNS) scale (Gagné, 2003; Ryan et al., 2006; Ryan & Deci, 2000) is a widely used, strongly validated measure of need satisfaction. One study looking at online and face-to-face learning contexts found that some items on the BPNS (in this instance relatedness measures) may need modifying in order to be appropriate for use online (Wang et al., 2019). It is therefore important that research into PWB online should consider the contextual appropriateness of BPNS and a more specific version of this scale which reflects the online domain is needed. The OWS will aim to fulfil this requirement.

SWB focuses on the area of life satisfaction in terms of positive and negative emotions or affect. SWB and PWB are separate constructs, but are very closely related and experienced together; people with high levels of both SWB and PWB can be categorised as 'flourishing' (Heintzelman, 2018) Benefits to SWB tend to be greater than PWB immediately after an experience but PWB benefits are higher in long term; activities that increase PWB often also increase SWB but vice versa is not necessarily the case (Heintzelman, 2018). Therefore measuring both is of value to understanding overall wellbeing. Satisfaction of BPN have also been repeatedly found to be positively associated with life satisfaction (Diener et al., 2017). Measures such as the Positive and Negative Affect Schedule (PANAS, Watson et al., 1988) and the Scale of Positive and Negative Experience (SPANE, Diener et al., 2010) are widely used to measure SWB. There is little work on the effect of either being online or using algorithmically mediated systems, although the conscious choice to use an algorithm has been found to reduce positive mood (Alexander et al 2018).

The basic psychological needs of SDT/PWB can be both experiential outcomes of *activities* and *motives* that directly influence behaviour. Different activities have been shown to satisfy the basic needs to greater or lesser extent; Martela & Sheldon (2019) suggest motives and activities should be measured alongside both SWB and PWB to get a rounded measure of wellbeing: "SWB only answers the question of *how* the subject is feeling, but not the question of *why* the subject is feeling so, or *what* he or she is doing" (p.7). It is likely that the many different activities people carry out online, from shopping for a new television online to interacting with friends on social media will have different effects. The current overabundant flow of online information and social relationships can easily contribute to users not being able to avoid excessive multi-tasking or overconsumption of media. Considering the context of a person's online life can speak to their motivations for being online, and relating these to wellbeing can help to identify which activities may be eudaimonic in nature or contribute to SWB.

Additionally, user understanding and digital literacy may also affect levels of online wellbeing (Gui et al., 2017). Users' digital skills (i.e., operational, technical and formal, information/cognition, digital communication, digital content creation and strategic skills) can influence users' online experiences and ability to cope with the side effects of over engagement, lack of transparency and agency experienced by many when online (Iordache et al., 2017). In keeping with the use of self-report measures, it may be appropriate to think of digital literacy as a measure of the users' perceived ability to navigate and stay safe in the online world. For example, a user with a high level of confidence in their ability may feel an increased sense of autonomy and competence, and lead to more positive experiences; however equally such a user may feel less autonomy due to the nature of online decision-making algorithms, and experience greater stress or frustration during particular activities.

## 2.2. The relationships between online wellbeing and trust

As noted previously, the increased use of automated decision-making algorithms online may lead to users being unsure whether they can trust the platforms that use them, which can lead to concerns about wellbeing, especially amongst potentially vulnerable users (Creswick et al., 2019; Knowles & Hanson, 2018). Therefore, alongside the OWS, a second scale measuring user trust online ('Trust Index' or TI) is also being developed, which can be used either concurrently or as a separate measure. The first stage of TI development involves identifying particular online factors that affect user trust, as well as identifying common opinions and experiences of trust that can then be transformed into a series of statements which will be validated in future studies. Outside of specific online contexts such as e-commerce and health information, work on trust in online platforms is relatively scarce. Bhattacherjee (2002) developed a scale for individual trust in online firms, for example Amazon, but there are no existing validated measures of online trust that can be adapted to take into account the general contexts in which trust is enacted. Trust is often measured by single statements or binary, yes/no, type questions, but it has been suggested it should be conceived as a complex and multidimensional psychological state, with both affective and motivational aspects (Kramer, 1999).

Trust is often considered in two forms: interpersonal and institutional. Considering algorithmically-mediated platforms, it appears that theories of institutional trust may be most appropriate, however some aspects of interpersonal trust may come into play if for example a user treats the website as a 'person', anthropomorphising the system. It has been found that people use similar neurophysiological mechanisms to trust algorithms as they would people, with companies taking advantage of this by 'personalising' decision aids (for example Apple's Siri or Amazon's Alexa) (Alexander et al., 2018). It has also been found that people who believed that others are generally trustworthy were more than twice as likely to adopt an algorithm (Alexander et al., 2018).

The most relevant existing work on trust in this context come from the fields of organisational psychology, management, and marketing. Within such research there is a broad range of conceptualisations, antecedents and types of trust (Kramer, 1999). Often research highlights antecedents of trust such as familiarity and reputation, security and other situational assurances, and encouraging 'normal' user experiences (Gefen et al., 2003, 2008; Yoon, 2002). Differences in levels of trust and related factors have been examined cross-culturally, for example between Japan and the US (Yamagishi & Yamagishi, 1994), indicating that it is also important to consider the different perceptions about trust that users may have, and the differing levels of importance they may place on trust, rather than sticking to a rigid definition and assumptions. As such, self-report is once again of value.

It is likely that someone's trust in a situation will affect their wellbeing, but if for example, trust is not a consideration for them when they are online, their wellbeing and behaviour may be less affected than someone for whom trust is highly important. Trust has been found to be a consistent predictor of SWB (Helliwell & Wang, 2011) although in Serbia this was only the case for institutional trust (Jovanović, 2016). Both interpersonal and institutional trust have been positively associated with life satisfaction in many different countries (Elgar et al., 2011). Recently, trust has also been linked to SDT, suggesting that the satisfaction of basic psychological needs can lead to a motivation to trust, and this motivation will affect people's willingness to continue to trust someone, or to restore trust (van der Werff et al., 2019). Measuring aspects of behaviour and wellbeing alongside trust may also lead to understanding of why and how trust in online systems manifests; it may also help to explain why and when people deviate from the expected trusting behavior, i.e., trust when they objectively should not, and vice versa, which they often do (McKnight et al., 1998).

## 3. METHOD

The first stage of development of the OWS and TI took place as part of a larger study into online trust, comparing attitudes of younger (16-25 years old) and older (over 65) adults. The study was approved by the Ethics Review Board at all co-authors' institutions as a joint research effort.

### 3.1. Participants

In total, 74 participants took part in nine 3-hour workshops between April and July of 2019; 5 workshops with 40 older adults (mean age 71 years, 62.5% female) and 4 workshops with 34 younger adults (mean age 20, 58.8% female). Recruitment took place through social media, fliers, and emails to groups such as University of the Third Age. Sessions were conducted in easily accessible venues in Nottingham and Oxford. Prior to the study participants were asked to confirm that they "regularly use the internet for searching for information, making bookings, or buying products" (89.2% indicated they used the internet several times a day) and were thanked afterwards with a £20 high street voucher.

### 3.2. Design

The project focused on user-driven, human-centred, and Responsible Research and Innovation (Jirotka et al., 2017) approaches to investigating trust. Thus the workshop structure, including timings and ordering of activities, the tasks themselves, and practical considerations were co-created through a series of activities with members of the public in the relevant age groups, ensuring that the content was relevant, understandable, and engaging. The final workshop structure, content and duration was decided through this iterative, user-centred piloting process. The resulting workshops took a mixed-methods approach to encourage participants to think about issues in different contexts. As such the workshops consisted of four distinct activities, with all participants taking part in each activity: (1) a quasi-naturalistic experiment observing user behaviour online, involving a screen-based task in which participants were asked to carry out a common online task (booking a hotel) on two different sites; (2) scenario-based discussions of trust; (3) a paper-based group task looking at different ways of presenting information to identify user requirements for an online tool to negotiate trust; and (4) pre- and post- session questionnaires measuring wellbeing related to trust online. As such, each activity can stand alone or be combined to compare responses in different contexts.

### 3.3. Materials and Procedure

This paper focuses on the results of the questionnaire activity. The questionnaires were designed to explore whether there is a link between trust, motivation, digital literacy, and wellbeing factors, and how this might be measured. They consisted of a mixture of free text, multiple choice, and Likert-like items. A pre-session questionnaire asked about: *Online Activity:* how often people go online, plus the primary reason and any secondary reasons that they go online out of 5 activities: socialising, buying or booking things, finding information, watching videos or playing games, and sharing content; *Trust:* rating of the importance of trust, whether they have ever stopped using a site because of a lack of trust, and 4 trust-related statements on 7-point Likert-like scales; and *Digital Confidence:* 7 statements on 7-point Likert-like scales, related to perceived digital literacy and how confident users are in carrying out tasks online.

A post-session questionnaire began with questions about the session, then repeated statements from the pre-session questionnaire to see if there were changes in opinion, followed by: *Trust:* ratings on a 7-point Likert-like scales of how much 12 different features of websites affect their trust, and an open-ended question about which factor is most important; and *Wellbeing:* an open-text question about whether the online world affects wellbeing, plus 2 instruments for measuring wellbeing, modified to reflect online experiences. PWB was measured using the Basic Psychological Need Satisfaction (BPNS) scale (Gagné, 2003; Ryan et al., 2006; Ryan & Deci, 2000), including 18 statements on a 7-point Likert-like scale. The scale was minimally modified to reflect the online world, for example "I feel like I am free to decide for myself how to live my life" was altered to "I feel like I am free to decide for myself how to act online", whilst other statements simply had the word 'online' added for context. SWB measurement used a format similar to the Scale of Positive and Negative experience (SPANE) (Diener et al., 2010), including 8 positive and 8 negative feelings on a 7-point Likert-like scale. The measure included words that are often used by Internet users to describe their experiences (Creswick et al., 2019). As suggested by Diener et al. (2010) the amount of time the state was felt rather than intensity of feeling was captured as it is stronger with regards to life satisfaction. A 7-point scale was chosen throughout in order to be consistent with the BPNS, as the most widely validated set of statements.

### 3.4. Analysis

The analysis of the quantitative data reported in this paper was carried out in SPSS. Cronbach's alpha reliability analysis was carried out on the scale items to check the internal consistency of items. In general, an alpha of 0.7 to 0.8 is deemed acceptable, with 0.8 more appropriate for cognitive measures, although at feasibility phase, scores as low as 0.5 may be accepted (Field, 2013). The results make no assumptions about the unidimensionality of the scales at this stage. Qualitative analysis of questionnaire responses was carried out using an inductive, data driven approach, to identify themes that were strongly interrelated with the raw data (Braun & Clarke, 2006). A single researcher analysed and coded all the responses fully using NVivo. Another researcher checked this coding for consistency and agreement, and any discrepancies were discussed and resolved. The codes were then grouped into key themes.

## 4. RESULTS AND DISCUSSION

The results here look at the relevant sections of the questionnaires for development of the prototype OWS and TI. Table 1 summarises each subscale that will be included in the OWS and its properties, based on the following results. The decision to change from a 7 to a 5-point scale for all subscales was taken for consistency with the new wellbeing measures, which are discussed below.

Table 1. Online Wellbeing Scale subscales, Trust Index subscales, and properties.

| OWS Subscale | Items | Score range |
|---|---|---|
| Online Activity | 6 | 6 (very little activity) to 30 (a lot of activity) |
| Digital Confidence | 6 | 6 (very low confidence) to 30 (very high confidence) |
| Psychological Wellbeing | 18 | |
| - Autonomy Dis/Satisfaction | - 3/3 | 3 (very low need dis/satisfaction) to 15 (very high need dis/satisfaction) |
| - Competence Dis/Satisfaction | - 3/3 | |
| - Relatedness Dis/Satisfaction | - 3/3 | |
| - Overall Dis/Satisfaction | - 9/9 | -12 (max. dissatisfaction) to 12 (max. satisfaction) |
| Subjective Wellbeing | 12 | |
| - Positive/Negative Affect | - 6/6 | 6 (very low +/- affect) to 30 (very high +/- affect) |
| - Affect Balance | 12 | -24 (unhappiest possible) to 24 (happiest possible) |
| **TI Subscale** | | |
| Importance | 6 | 6 (not at all important) to 30 (highly important) |
| Belief | 6 | 6 (not at all trusting) to 30 (highly trusting) |
| Context | 16 | For each item, 1 (not important to trust) to 5 (very important to trust) or 1 (not trustworthy) to 5 (very trustworthy) |

### 4.1. The Online Wellbeing Scale

Overall, 66.2% of participants felt that the online world and their use of the internet affected their wellbeing, and of these 63.8% felt that its effect was negative, with a further 8.5% suggesting that it could be both. This shows the importance of developing ways to measure the effects of the online world on wellbeing. Participants did a wide range of online activities, with an average of 4 of the 5 options being selected; all activities were chosen by at least half of the participants. The most common activities were finding information (95.9%), socialising (85.1%) and buying or booking things (82.4%). The most common primary activity was socialising (35.1%) followed by finding information (31.1%). Other suggestions for online activities over the last four weeks were made by 18.9% showing that the original list covered most common online activities. The other suggestions were email (8.1%), banking (8.1%) and work (2.7%). This has contributed to the *Online Activity* block of the OWS, containing the items from the initial questionnaire, plus an additional item, 'financial or organisation' which covers all activities suggested by participants. This 6-item block measures the frequency of each activity in an additive way, from 1 (very rarely or never) to 5 (very often or always) for each activity. This allows for a measure of both the range and extent of online activity.

The 7 statements which related to digital confidence from the pre-session questionnaire have cronbach's alpha (α) reliability of 0.797; removing one item ("If I do not trust a website I can do something about it") improves reliability to α=0.827. Modifying one statement slightly from "I am able to tell whether a website is trustworthy" to "Your ability to tell whether or not a website is trustworthy", the *Digital Confidence* block contains 6 items to be rated from 1 (very low) to 5 (very high).

The BPNS from the post-session questionnaire had low reliability for autonomy (α=0.581) and competence (α=0.454). Only relatedness reached an acceptable level (α=0.792). As such, the scale as a whole is not considered reliable. Removing items increased reliability for each scale (α=0.636, α=0.515, α=0.812 respectively) but not enough for the whole scale to be considered acceptable. Many participants also skipped items, for example only 53 participants completed the relatedness items. It was noted that, particularly for the older participants, statements relating to interacting with people online were often either ignored or misunderstood. For the prototype OWS therefore, the *Psychological wellbeing* scale will be replaced with a modified version of the Balanced Measure of Psychological Needs (Sheldon & Hilpert, 2012). This scale uses simpler language and reduces each construct to 6 items, with the ability to calculate the overall level of satisfaction and dissatisfaction of needs. It is also easily modified for different domains. Modifications to this scale will include focusing wording on the online world and replacing specific references to 'people' with a more general interactional focus.

The SWB measure scored acceptable reliability for the positive experience scale of α=0.775, improved by removing one item ("High Mood") to α=0.785, and the negative experience scale scored a good reliability of α=0.805, improved by removing one item ("Tracked") to α=0.808. As each state was presented with an opposite state, removing the equivalent positive and negative words ("Safe" and "Low Mood") resulted in a reliability of α=0.797 for positive experience and α=0.781 for negative experience. This results in a modified *Subjective Wellbeing* scale of 6 items each for positive and negative feelings: Calm/Anxious; Creative/Apathetic; Empowered/Disempowered; Pleased/Annoyed; Powerful/Powerless. This scale is used to derive an overall affect balance score and can also be divided into positive and negative feelings scales.

**4.2. The Trust Index**

The Trust statements from the pre-session questionnaire do not form a coherent scale, with the highest cronbach's alpha score reaching just 0.249. It was also found that there are several different aspects to trust that apply to the online world. As such, in order to create the TI prototype, three different aspects of trust will be measures in separate subscales. Table 2 summarises the subscales and properties. The *Importance* of trust will form an individual score for how important trust is to the individual when they are online. A total of 78.4% of participants had left a website or stopped using an online service because they did not trust it, suggesting that to some trust is a highly important factor when they are online. 50% of participants agreed that it was indeed highly important, but 9.5% felt that it was not very important at all, so to those participants, trust may not affect their online wellbeing or behavior. Additionally, 83.8% agreed that "websites have a responsibility to act in a trustworthy manner towards their users". Therefore, *Importance* is worth measuring in this scale.

The trusting *Beliefs* of an individual will form an overall score for how much that individual trusts or believes in the online services they use in general. Responses to "I tend to trust things I find on the internet" were varied, with 40.6% disagreeing, 28.4% agreeing, and the remainder responding neutrally. Similarly, 32.5% agreed that "websites do enough to make sure their users trust them" with 25.7% disagreeing. If a person has a general lack of trust in the online world, this may affect their behaviour and wellbeing in different ways to if a person is generally trusting of what they do and see online (Alexander et al., 2018; Elgar et al., 2011; Helliwell & Wang, 2011). Combined with their feelings about the importance of trust this may give a good indication of the levels of trust a user is likely to have in an online service or platform. A 'general trust scale' (Kramer, 1999; Yamagishi & Yamagishi, 1994), examining beliefs about trust and honesty with regards to other people will be modified to relate to online platforms.

The remaining subscale will relate to different *Contexts* and factors related to going online and whether these affect levels of trust, plus a final item which measures whether an individual thinks that their online activities influence their sense of wellbeing. This is to allow a quick comparison of trusting beliefs with their feelings about wellbeing, to compliment the OWS. The contextual items include online reviews, financial transactions, social media, recommendations and entertainment, as the main activities that participants flagged as important, and relating to those they indicated taking part in most often. All of the factors presented to participants in the questionnaire were deemed important, with the most important being reputation ($\bar{x}$=6.11) and use of personal data ($\bar{x}$=5.99). Items related to ease-of use also scored highly ($\bar{x}$=5.50). These also matched with the qualitative responses asking for the most important factors: *"well known firm, been there before"; "the company and its reputation and previous knowledge of it"; "the information which it is demanding for me to share with it, and the language which is used surrounding this"; "difficult to navigate"; "user friendly structure"*. The factor that emerged the most in qualitative responses was security measures: *"the padlock item in the top to let me know it is secure"; "checking whether it's a secure site".* Familiarity with the site was also important: *"if I have used it before"; "whether I know about it/I have had previous experience using it"*. The other factors that are included therefore are: brand and social reputation, personal data, reliability and ease of use, familiarity with the website, privacy policies and security measures, and algorithms. Several of these also relate strongly to factors or antecedents found in the literature on trust, for example familiarity, brand reputation, security, and reliability (Gefen et al., 2003, 2008; Kramer, 1999; Yoon, 2002); the reliance on familiarity and reputation also corresponds to the finding that social proof (i.e., knowing that other people has used it) was found to be the most effective way to persuade people to adopt algorithms (Alexander et al., 2018). This subscale will not be scored like the others, but will be used to identify areas where trust is most relevant or has most effect on users' experiences.

## 5. CONCLUSIONS AND NEXT STEPS

The questionnaires completed during a series of workshops with younger and older adults allowed for preliminary examination of how online trust and wellbeing might usefully be measured, in order to develop an Online Wellbeing Scale and Trust Index. Examination of the internal reliability of several groups of statements aimed at measuring different factors has led to the development of prototype scales. The 42-item OWS has 4 subscales, with each construct measured by a series of statements on 5-point Likert or Likert-like scales. The first two blocks provide a baseline for the users' online experience: *Online activity* examines the range and amount of activity the participant carries out online whilst *Digital confidence* forms a self-report measure of digital literacy. Both of these factors are likely to have an effect on both online wellbeing and trust. The second 2 blocks examine two major conceptualisations of wellbeing: *Psychological* (eudaimonic) *wellbeing* and *Subjective (*hedonic) *wellbeing.* The TI, which can be used standalone or in cohort with the OWS, consists of 29 statements covering 3 subscales measured by a 5-point Likert Scale. These cover three areas of trust which emerged as important to trust in the workshops and the questionnaire results, with reference to related literature on trust: the *Importance* of trust to the individual user*,* the *Belief* of the user whether or not the online world is trustworthy*,* and the *Context* in which trust or distrust is enacted.

The modified prototypes of the OWS and TI are currently being tested in a large online study. This will allow both validation of the scales and large-scale examination of the role of trust in online wellbeing. As well as testing the internal reliability of the subscales, a principal components analysis will be carried out with other calculations of dimensionality to ensure that the subscales measure individual factors within trust and wellbeing. This will be especially important in considering the *Context* subscale of the TI. The results of the online study will also help lead to recommendations for ways in which online platforms can build user trust into their systems. At the same time, using the TI for reflection and empowerment will be explored with users and other stakeholders, including investigating ways to present results that are meaningful and engaging, and how this and other tools can encourage meaningful dialogue between the two groups.

**ACKNOWLEDGEMENTS**

This work was supported by EPSRC [grant number EP/R033633/1]. Elvira Perez Vallejos also acknowledges the financial support of the NIHR Nottingham Biomedical Research Centre.

DEVELOPING A MEASURE OF ONLINE WELLBEING AND USER TRUST